\documentclass[twocolumn,showpacs,preprintnumbers,amsmath,amssymb,endfloats*]{revtex4}
\usepackage{graphicx}

\begin{document}
\def \ee {\varepsilon}
\thispagestyle{empty}
\title{
Conductivity of dielectric and thermal atom-wall interaction
}

\author{
G.~L.~Klimchitskaya\footnote{on leave from North-West Technical University,
 St.Petersburg, Russia}
and V.~M.~Mostepanenko\footnote{on leave from Noncommercial Partnership
``Scientific Instruments'',  Moscow,  Russia}
}

\affiliation{
Center of Theoretical Studies and Institute for Theoretical
Physics, Leipzig University,
D-04009, Leipzig, Germany
}

\begin{abstract}
We compare the experimental data of the first
measurement of a temperature dependence of the
Casimir-Polder force by Obrecht et al. [Phys.
Rev. Lett. {\bf 98}, 063201 (2007)] with the
theory taking into account small, but physically
real, static conductivity of the dielectric
substrate. The theory is found to be inconsistent
with the data. The conclusion is drawn that the
conductivity of dielectric materials should not
be included in the model of the dielectric
response in the Lifshitz theory. This conclusion
obtained from the long separation measurement is
consistent with related but different results
obtained for semiconductors and metals at short
separations.
\end{abstract}

\maketitle

The Casimir-Polder force \cite{1} acts between rarefied atoms and a
wall. It originates from zero-point and thermal fluctuations
of the electromagnetic field.
At separations $x$ less than a few nanometers (but larger than a few
angstroms) the interaction between an atom and a wall is of nonrelativistic
character and takes the form of the nonretarded van der Waals force.
The interaction potential in this region, $V_3(x)=-C_3/x^3$, was found
by Lennard-Jones \cite{2}. Casimir and Polder included
retardation effects which result in the interaction
potential $V_4(x)=-C_4/x^4$ at separations of about $1\,\mu$m. In the
intermediate region the quantitative description of the
van der Waals-Casimir-Polder force invites for the inclusion of material
properties. Thermal effects come into play at separations larger than
about $2\,\mu$m. Both material properties and thermal effects were
included in the Lifshitz theory of the van der Waals forces between
dielectrics \cite{3}. This theory describes the case of two parallel
plates (separated by a vacuum gap of width $d$) and
that of an atom at a separation
$d$ from a wall.

For a long time the study of atom-wall interaction was considered as merely an
academic exercise because the interaction is relatively small. The
situation changed radically during the last decade when the atom-wall
interaction in different physical, chemical and biological processes
received increasing attention \cite{4}. As examples,
this interaction plays major role in experimental studies of quantum
reflection, Bose-Einstein condensation and diffraction of atoms on
different surfaces \cite{5,6,7,8,9,10,10a}. It attracts additional interest
in nanotechnological applications, such as carbon nanotubes \cite{11,12},
and mesoscopic-scale atomic devices. All these applications require accurate
characterization of atom-wall interaction including the dependence of the
force on atomic and material properties and on the temperature (see
\cite{13,14} for the most precise computations on the basis of the
Lifshitz theory and for a history of the problem). In parallel with the
Casimir-Polder interaction, the Casimir force between two macrobodies
\cite{15} was investigated both experimentally and theoretically
(see reviews \cite{16,17} and recent experiments \cite{18,19}).
Similar to the Casimir-Polder force, the Casimir force finds
multidisciplinary applications ranging from constraints on hypothetical
interactions predicted by multi-dimensional physics \cite{18}
to nanotechnological applications \cite{20}.
Comparison of the
Casimir force measurements with theory also needs to be done with
inclusion
of real material properties and nonzero temperature \cite{17}.
The temperature dependence of the Casimir force between two metal plates is
the subject of discussion (see, e.g., \cite{21,22}).
Some of the measurements of the Casimir force between metallic macrobodies
performed up to date \cite{18} were precise enough
to exclude theoretical models which take into account
the relaxation processes in the real current of conduction electrons.
Until now, however, experiments were not of sufficient precision
to measure the thermal effect in situ.

The first measurement of the thermal Casimir-Polder force was performed in
the excellent experiment \cite{23} both in thermal equilibrium and
in the nonequilibrium case. In that experiment the dipole
oscillations with the frequency $\omega_0$ were excited in a ${}^{87}$Rb
Bose-Einstein condensate separated by a distance of a few micrometers
from a fused-silica substrate (wall). The Casimir-Polder force $F_{CP}$
between a rubidium atom and a substrate changes the magnitude of the
oscillation frequency making it equal to some $\omega_x$. The fractional
frequency difference $\gamma_x=|\omega_0-\omega_x|/\omega_0$ was
measured and compared with theory \cite{24,27} at an environment temperature
$T_E=310\,$K and at different substrate temperatures $T_S=310\,$K
(thermal equilibrium) and 479\,K, 605\,K (out of thermal equilibrium).
In all cases excellent agreement between data and theory
was demonstrated.

In this paper, we obtain important new information from the
measurement data of experiment \cite{23}. Using theory developed in
Refs.~\cite{24,27},
we recalculate the fractional frequency difference $\gamma_x$
by taking into account a nonzero static conductivity of fused silica at
experimental temperatures.
We emphasize that introduction of
such a small but real conductivity will lead to a drastically
different result in the framework of the Lifshitz theory.
The obtained theoretical results both in
thermal equilibrium and out of thermal equilibrium are in {\it disagreement}
with the measurement data. At the same time, when we neglect the static
conductivity
of fused silica in our calculations, we return back to the computational
results of Ref.~\cite{23} which are in excellent agreement with the data.
We have checked that the differences between the theoretical results computed
at different $T_S$ with the static conductivity of fused
silica included and neglected are completely determined by the
equilibrium contribution $F_{CP}$ to the total atom-wall force
$F(x,T_S,T_E)$ [see Eq.~(\ref{eq3}) below].
The additional contributions
$F_n(x,T_S)$ and $F_n(x,T_E)$
arising in the atom-wall force
out of thermal equilibrium are the same, regardless of whether
the static conductivity of the substrate material is included.
The obtained experimental evidence that the static conductivity of a
wall should be neglected in theoretical calculations using the
Lifshitz theory is of much interest for the numerous applications of the
Casimir-Polder and Casimir forces. It amplifies the measurement results
of Ref.~\cite{19}, demonstrating that to achieve an agreement between
experiment and theory in the Casimir interaction of a gold sphere and
a semiconductor plate with doping concentartion below critical, the
static conductivity of semiconductor should be neglected.
Another type of experiments \cite{18,25a,25,26} deals with the Casimir
interaction between two gold macrobodies. These experiments also
demonstrate that theory taking into account the relaxation processes in
real current of conduction electrons is inconsistent with measurement
data. To achieve an agreement between data and theory, one should neglect
the real current of conduction electrons.

As opposed to \cite{18,19,25a,25,26}, the above evidence is based on the
measurement data related to a dielectric sample. It is the first one for
any kind of material which
follows from the data at separation distances
of a truly thermal regime above $6\,\mu$m (experiments
in Refs.~\cite{18,19,25a,25,26} lead to conclusive
results at separations of $1\,\mu$m and less).

In the experiment \cite{23} the unperturbed trap frequency was measured to be
$\omega_0=2\pi\times229\,$Hz. The separation distance between the trap center
of mass and the surface of the substrate $d$ varies from 6.5 to $11\,\mu$m.
Now we briefly present the calculation procedure of the perturbed
oscillation frequency $\omega_x$, as developed in Refs.~\cite{24,27}
(see also \cite{29,30} for some details).

In accordance with the Lifshits theory, the Casimir-Polder force between
an atom at a separation $x$ above a substrate at a temperature
$T$ in thermal
equilibrium is given by
\begin{eqnarray}
&&
F_{CP}(x,T)=-2k_BT\left[
\vphantom{\sum_{l=1}^{\infty}}
\alpha_0r_0\int_{0}^{\infty}\!\!\!k^3
dk{\rm e}^{-2kx}\right.
\nonumber \\
&&~~~~
+
\left.\sum_{l=1}^{\infty}\alpha_l
\int_{0}^{\infty}\!\!\!k
dk{\rm e}^{-2q_lx}h(\xi_l,k)\right],
\label{eq1} \\
&&
h(\xi_l,k)=\left(2q_l^2-\frac{\xi_l^2}{c^2}\right)
r_{\rm TM}({\rm i}\xi_l,k)-\frac{\xi_l^2}{c^2}
r_{\rm TE}({\rm i}\xi_l,k).
\nonumber
\end{eqnarray}
\noindent
Here, the dielectric permittivity of the substrate
$\varepsilon_l=\varepsilon({\rm i}\xi_l)$ and the atomic dynamic
polarizability $\alpha_l=\alpha({\rm i}\xi_l)$ are calculated at the
imaginary Matsubara frequencies, $\xi_l=2\pi k_BTl/\hbar$, and the
following notations are introduced
\begin{eqnarray}
&&
r_{\rm TM}=\frac{\varepsilon_lq_l-k_l}{\varepsilon_lq_l+k_l},
\quad
r_{\rm TE}=\frac{q_l-k_l}{q_l+k_l}, \quad
r_0=r_{\rm TM}(0,k),
\nonumber \\
&&
q_l=\sqrt{k^2+\frac{\xi_l^2}{c^2}}, \qquad
k_l=\sqrt{k^2+\varepsilon_l\frac{\xi_l^2}{c^2}}.
\label{eq2}
\end{eqnarray}

When the temperature of the substrate ($T_S$) and of the surrounding
environment ($T_E$) are different, the force acting between an atom
and a substrate was obtained in Ref.~\cite{27}
\begin{equation}
F(x,T_S,T_E)=F_{CP}(x,T_E)+F_n(x,T_S)-F_n(x,T_E),
\label{eq3}
\end{equation}
\noindent
where the nonequilibrium contribution is defined as
\begin{eqnarray}
&&
F_n(x,T)=-K
\int_{0}^{\infty}\!\!d\omega
\int_{0}^{\infty}\!\!\!dtf(\omega,t){\rm e}^{-\frac{2\omega tx}{c}},
\label{eq4} \\
&&
f(\omega,t)=\frac{\omega^4t^2}{{\rm e}^{\frac{\hbar\omega}{k_BT}}-1}
\left[|p(\omega,t)|+
{\rm Re}\varepsilon(\omega)-1-t^2\right]^{1/2}
\nonumber \\
&&~~
\times\left[\frac{1}{|\sqrt{p(\omega,t)}+{\rm i}t|^2}
+\frac{(2t^2+1)(t^2+1+
|p(\omega,t)|}{|\sqrt{p(\omega,t)}+
{\rm i}\varepsilon(\omega)t|^2}\right],
\nonumber \\
&&
K=\frac{2\sqrt{2}\hbar\alpha_0}{\pi c^4}, \qquad
p(\omega,t)\equiv\varepsilon(\omega)-1-t^2.
\nonumber
\end{eqnarray}

The frequency shift of the condensate oscillations under the influence of
the force $F$ is obtained as \cite{27}
\begin{eqnarray}
\omega_0^2-\omega_x^2&=&-\frac{\omega_0}{\pi am}\int_{0}^{2\pi/\omega_0}
\!\!\!\!\!d\tau\cos(\omega_0\tau)
\label{eq5} \\
&&\times
\int_{-R_x}^{R_x}\!\!\!d\tilde{x}
n_0^{x}(\tilde{x})F[d+\tilde{x}+a\cos(\omega_0\tau),T_S,T_E].
\nonumber
\end{eqnarray}
\noindent
In the experiment \cite{23} $a=2.50\,\mu$m is the amplitude of the
oscillations, $R_x=2.69\,\mu$m is the Thomas-Fermi radius in the
$x$-direction, $m=1.443\times 10^{-25}\,$kg is the mass or rubidium atom
and
\begin{equation}
n_0^x(\tilde{x})=\frac{15}{16R_x}\left(1-\frac{\tilde{x}^2}{R_x^2}
\right)^2.
\label{eq6}
\end{equation}

 As a supplementary element to Refs.~\cite{24,27,29,30},
we perform analytically the averaging procedures.
Substituting Eqs.~(\ref{eq3}) and (\ref{eq6}) into Eq.~(\ref{eq5}) and
integrating with respect to $\tilde{x}$ and $\tau$, one
arrives at
\begin{equation}
\gamma_x=\frac{1}{ma\omega_0^2}\left|\Phi_{e}(d,T_E)+
\Phi_{n}(d,T_S)-\Phi_{n}(d,T_E)\right|,
\label{eq7}
\end{equation}
\noindent
where
\begin{eqnarray}
&&
\Phi_{e}(d,T)=-2k_BT\left[
\vphantom{\sum_{l=1}^{\infty}}
\alpha_0r_0\!\int_{0}^{\infty}\!\!\!\!\!
k^3dk
{\rm e}^{-2kd}I_1(2ka)g(2kR_x)
\right.
\nonumber \\
&&\left.
+\sum_{l=1}^{\infty}\alpha_l\int_{0}^{\infty}
\!\!\!kdk
h(\xi_l,k){\rm e}^{-2q_ld}I_1(2q_la)g(2q_lR_x)\right],
\label{eq8} \\
&&
g(z)\equiv\frac{15}{z^5}[(3+z^2)\sinh{z}-3z\cosh{z}],
\nonumber
\end{eqnarray}
and $I_1(z)$ is the Bessel function. The function $\Phi_n(d,T)$ is
given by Eq.~(\ref{eq4}) for $F_n(x,T)$
where $x$ should be replaced with $d$ and the function $f(\omega,t)$ with
\begin{equation}
\tilde{f}(\omega,t)=f(\omega,t)
I_1\Bigl(\frac{2a\omega t}{c}\Bigr)g\Bigl(\frac{2R_x\omega t}{c}\Bigr).
\label{eq9}
\end{equation}

Now we are in a position to compute $\gamma_x$ under different
assumptions of the conductivity of fused silica.
Following Ref.~\cite{27}, the static dynamic polarizability of
Rb atoms $\alpha_l\approx\alpha_0=4.73\times 10^{-23}\,\mbox{cm}^{-3}$
is used in computations. This allows one to obtain highly accurate results
at separations under consideration \cite{13}.
For fused silica $\varepsilon({\rm i}\xi_l)$ as a function  of $\xi_l$
is taken from \cite{14}.

The computational results for $\gamma_x$ in thermal equilibrium
are shown in Fig.~1 by
the solid line. In the same figure, the experimental data
obtained in Ref.~\cite{23} at separations below $10\,\mu$m
are shown as crosses.
The absolute errors in the measurement of separations and $\gamma_x$
are presented in true scales at each individual data point.
The solid line in Fig.~1 practically coincides with the solid line
in Fig.~4(a) of  Ref.~\cite{23} computed using
$\varepsilon({\rm i}\xi_l)=\varepsilon_0=3.81$
(only minor deviations are observed at $d<8\,\mu$m). Note that in
the theory of Refs.~\cite{24,27,29} the frequency-dependent
$\varepsilon$ of fused silica was used with a finite static value
$\varepsilon_0$. It was shown that the account of the frequency
dependence leads to only a  minor effect and only at small
distances. We confirm this conclusion.
The theoretical computations are in excellent agreement with the data,
as was stated in Ref.~\cite{23}. Our computation
results for $\gamma_x$ in
nonequilibrium situations are presented by the solid lines in
Fig.~2(a) for $T_S=479\,$K and in Fig.~2(b) for $T_S=605\,$K. These
lines also practically coincide with respective lines in Fig.~4(a)
of Ref.~\cite{23}. According to \cite{27,29} the frequency dependence
of $\varepsilon(\omega)$ does not affect the nonequilibrium
contributions $\Phi_n$ in the studied ranges of $d$ and $T$.
 It is seen that at
nonequilibrium the data are also in a very good agreement with
theoretical computations.

At nonzero temperature any dielectric possesses nonzero static
conductivity. It can be taken into account by replacing the dielectric
permittivity $\varepsilon(\omega)$ with
\begin{equation}
\tilde\varepsilon(\omega)=\varepsilon(\omega)
+{\rm i}\frac{4\pi \sigma_0(T)}{\omega}.
\label{eq10}
\end{equation}
\noindent
The inclusion of conductivity dramatically affects the calculation
results using Eqs.~(\ref{eq7}), (\ref{eq8}). In the above computations
$r_0=(\varepsilon_0-1)/(\varepsilon_0+1)$ was used following
from Eq.~(\ref{eq2}) for $\varepsilon(0)=\varepsilon_0$.
The same Eq.~(\ref{eq2}) with the use of dielectric permittivity
(\ref{eq10}) results in $r_0=1$. This alone changes the
contribution from $\Phi_{e}(d,T)$ defined in Eq.~(\ref{eq8}) and
leads to the corresponding change in the magnitudes of $\gamma_x$ computed
using Eq.~(\ref{eq7}). We emphasize that this change does not depend on
the value of $\sigma_0$, but only on the fact that it is nonzero.
At $T_S=T_E=310\,$K the conductivity of fused silica sample varies within
a wide region from $10^{-9}\,\mbox{s}^{-1}$ to
$10^{2}\,\mbox{s}^{-1}$ depending on the concentration of impurities
\cite{32,33}. This results in negligibly small additions to
$\varepsilon_l=\varepsilon({\rm i}\xi_l)$ at all $\xi_l\neq 0$.

The computational results for $\gamma_x$ using Eqs.~(\ref{eq7}), (\ref{eq8})
and (\ref{eq10}) in thermal equilibrium are shown in Fig.~1 as the dashed
line. As is seen in the figure, the first two experimental points are in
clear disagreement with theory taking into account the conductivity
of fused silica.

In the nonequilibrium situation the disagreement between
the experimental data
and theory taking the static conductivity of silica into account
widens. Direct computations show that the addition of
conductivity does not influence the nonequilibrium contributions into
$\gamma_x$. The magnitudes of $\Phi_n(d,T)$
computed with different values of $\sigma_0$ from 0 to
$10^3\,\mbox{s}^{-1}$ coincide up to 6 significant figures.
Thus, the conductivity influences only through the equilibrium term
$\Phi_{e}(d,T)$. The respective results for $\gamma_x$ are presented
in Fig.~2 by the dashed lines. As is seen in Fig.~2(a) ($T_S=479\,$K),
the three experimental points exclude the dashed line and the other
two only touch it. The dashed line in Fig.~2(b) ($T_S=605\,$K)
demonstrates that all data exclude the theoretical prediction
incorporating the static conductivity of fused silica.
Thus, the confidence at which the theoretical approach based on
Eq.~(\ref{eq10}) is excluded by data increases with the increase
of substrate temperature $T_S$. This is in accordance with the
conclusion of Ref.~\cite{23} that the Casimir-Polder force for
a 605\,K substrate is nearly 3 times larger than for a 310\,K
substrate. The comparison of the complete set of data, as given by
crosses in Figs.~1 and 2(a,b), with the dashed lines shows
that the inclusion of the static conductivity of fused
silica in computations of the Casimir-Polder force is inconsistent with
the experimental data of Ref.~\cite{23}.

To conclude, we have
shown that the inclusion of a small, but physically
real, static
conductivity for the dielectric substrate in the
Lifshitz theory leads
to a large increase in the magnitude of the
Casimir-Polder force.
  However,
as is shown above,
the theoretical predictions including the static
conductivity of the dielectric
 are
in a deep disagreement with the experimental data of
Ref. \cite{23} for
 the
thermal Casimir-Polder force between an atom and a
silica surface.
Neglecting the static conductivity of
silica in the Lifshitz theory leads to an excellent
agreement with
the data. Thus, the static conductivity of dielectrics
should not be
included in the model of the dielectric response.
This conclusion is consistent with a
related but different experiment of the Casimir force
measurement
between a semiconductor plate and a gold sphere
\cite{19}, where it was
 found that
the static conductivity of the semiconductor plate
with doping
concentration below the critical should be neglected.
The obtained
results clarify the use of the Lifshitz theory and are
of wide
applicability in all multidisciplinary applications of
dispersion
forces.

The authors are greatly indebted to M.\ Antezza and R.\ J.\ Wild
for helpful discussions and providing details of calculation
procedure used in Ref.~\cite{23}. The discussions with G.\
Bimonte, C.\ Henkel
and U.\ Mohideen are also acknowledged. This work was supported by the
DFG Grant No.~436\,RUS\,113/789/0--3.


\widetext
\begin{figure}
\vspace*{-10cm}
\centerline{
\includegraphics{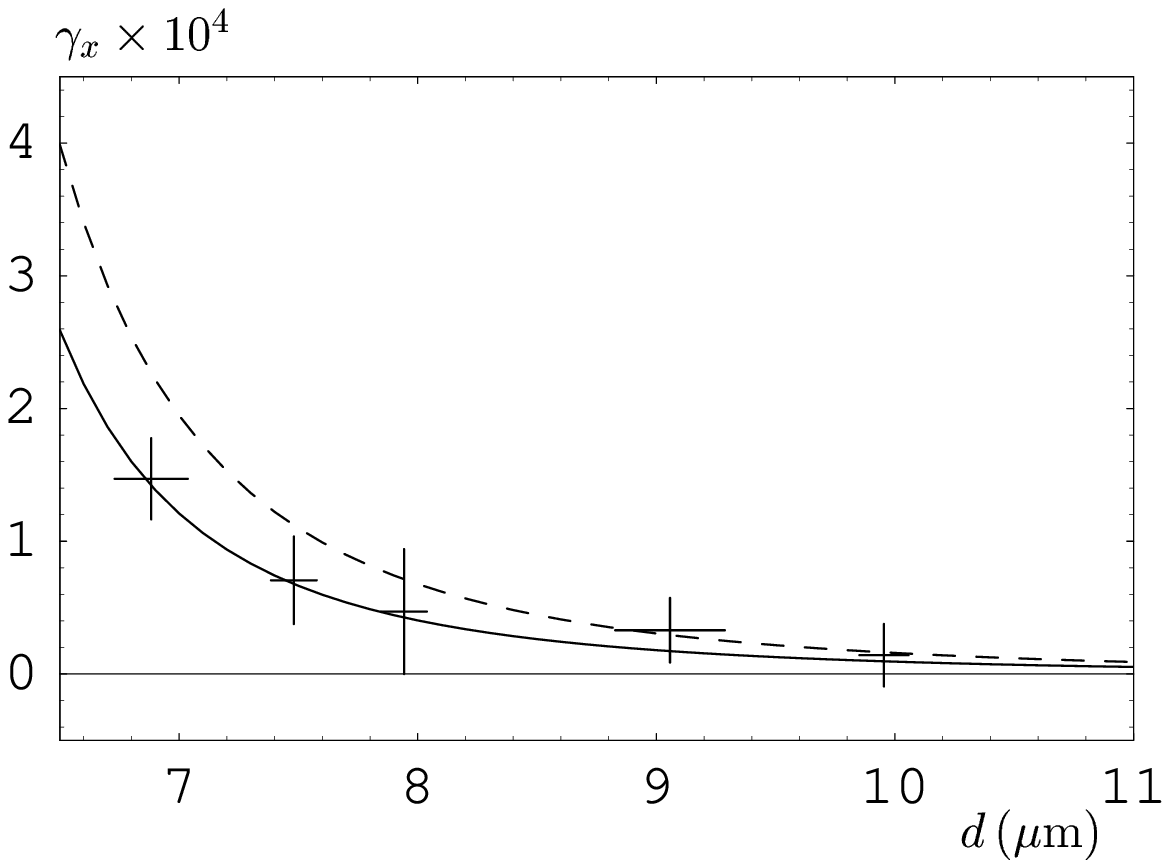}
}
\vspace*{-6cm}
\caption{
Fractional change in the trap frequency versus separation in thermal
equilibrium with $T_S=T_E=310\,$K computed by neglecting (solid line)
and including  (dashed line) the conductivity of the dielectric
substrate. The experimental data are shown as crosses.
}
\end{figure}
\begin{figure}
\vspace*{-5cm}
\centerline{
\includegraphics{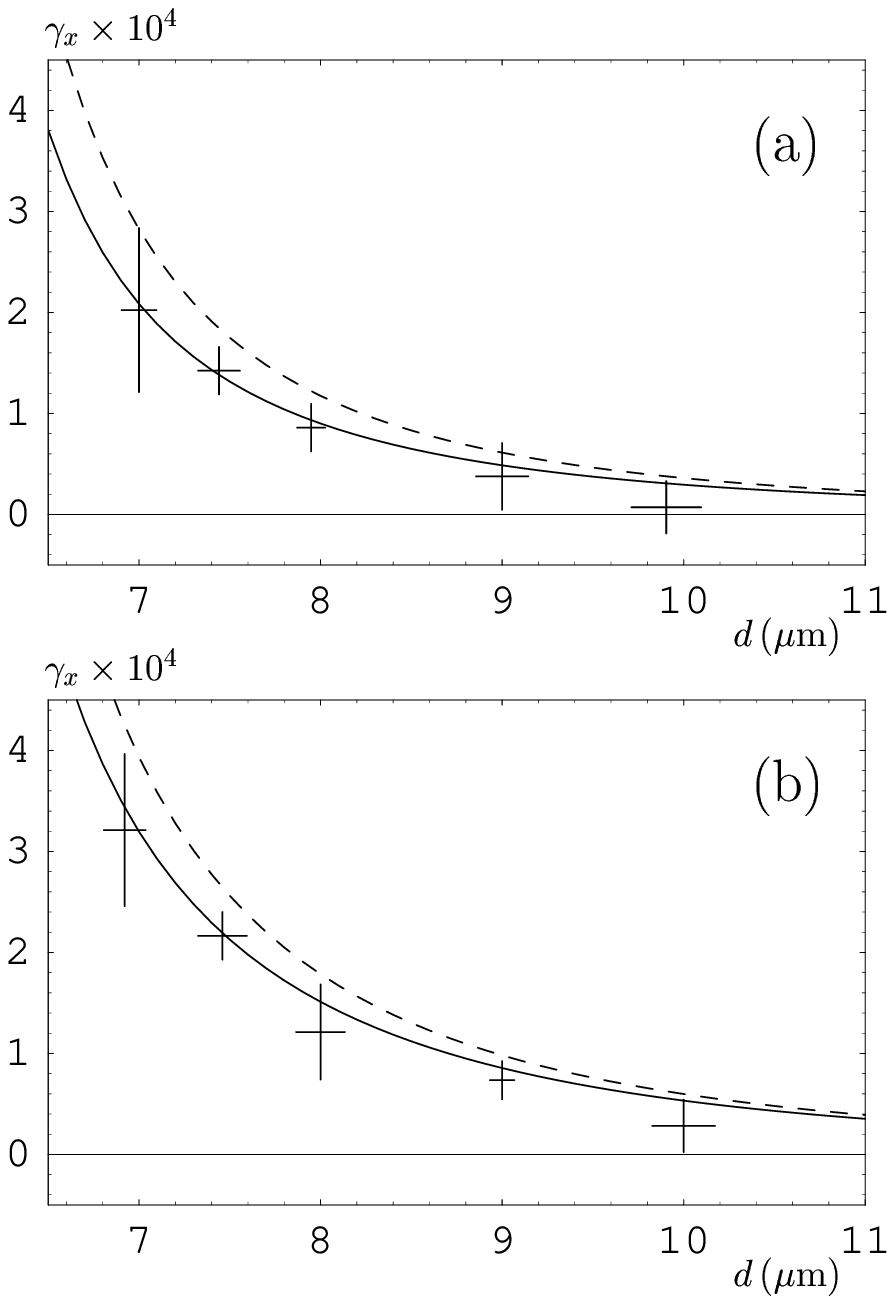}
}
\vspace*{-10cm}
\caption{
Fractional change in the trap frequency versus separation
out of  thermal equilibrium
(a) with $T_S=479\,$K and $T_E=310\,$K
and (b) $T_S=605\,$K, $T_E=310\,$K.
Computations are done by neglecting (solid line)
and including (dashed line) the conductivity of the dielectric
substrate. The experimental data are shown as crosses.
}
\end{figure}

\begin{thebibliography}{99}
\bibitem{1}
H.~B.~G.~Casimir and D.~Polder,
Phys. Rev. {\bf 73}, 360 (1948).
\bibitem{2}
J.~E.~Lennard-Jones,
Trans. Faraday Soc. {\bf 28}, 333 (1932).
\bibitem{3}
E.~M.~Lifshitz,
Zh. Eksp. Teor. Fiz. {\bf 29}, 94 (1956)
[Sov. Phys. JETP  {\bf 2}, 73 (1956)].
\bibitem{4}
V.~A.~Parsegian,
{\it Van der Waals Forces}
(Cambridge University Press, Cambridge, 2005).
\bibitem{5}
R.~E.~Grisenti, W.~Schollkopf, J.~P.~Toennies, G.~C.~Hegerfeldt,
and T.~Kohler,
Phys. Rev. Lett. {\bf 83}, 1755 (1999).
\bibitem{6}
F.~Shimizu,
Phys. Rev. Lett. {\bf 86}, 987 (2001).
\bibitem{7}
V.~Druzhinina and M.~DeKieviet,
Phys. Rev. Lett. {\bf 91}, 193202 (2003).
\bibitem{8}
Y.~Lin, I.~Teper, C.~Chin, and V.~Vuteli\'{c},
Phys. Rev. Lett. {\bf 92}, 050404 (2004).
\bibitem{9}
T.~A.~Pasquini, Y.~Shin, C.~Sanner, M.~Saba, A.~Schirotzek,
D.~E.~Pritchard, and W.~Ketterle,
Phys. Rev. Lett. {\bf 93}, 223201 (2004).
\bibitem{10}
H.~Oberst, Y.~Tashiro,  K.~Shimizu, and F.~Shimizu,
Phys. Rev. A {\bf 71}, 052901 (2005).
\bibitem{10a}
D.~M.~Harber, J.~M.~Obrecht, J.~M.~McGuirk,
and E.~A.~Cornell,
Phys. Rev. A {\bf 72}, 033610 (2005).
\bibitem{11}
E.~V.~Blagov, G.~L.~Klimchitskaya, and V.~M.~Mostepanenko,
Phys. Rev. B {\bf 71}, 235401 (2005);
ibid. {\bf 75}, 235413 (2007).
\bibitem{12}
M.~Bordag, B.~Geyer, G.~L.~Klimchitskaya, and V.~M.~Mostepanenko,
Phys. Rev. B {\bf 74}, 205431 (2006).
\bibitem{13}
J.~F.~Babb, G.~L.~Klimchitskaya, and V.~M.~Mostepanenko,
Phys. Rev. A {\bf 70}, 042901 (2004).
\bibitem{14}
A.~O.~Caride, G.~L.~Klimchitskaya, V.~M.~Mostepanenko,
and S.~I.~Zanette,
{Phys. Rev.} A {\bf 71} 042901 (2005).
\bibitem {15}
H.~B.~G.~Casimir,
{ Proc. K. Ned. Akad. Wet.}
{\bf 51}, 793 (1948).
\bibitem{16}
M.~Kardar and R.~Golestanian,
Rev. Mod. Phys. {\bf 71}, 1233 (1999).
\bibitem{17}
M.~Bordag, U.~Mohideen, and V.~M.~Mostepanenko,
{ Phys. Rep.} {\bf 353}, 1 (2001).
\bibitem{18}
R.~S.~Decca, D.~L\'opez, E.~Fischbach, G.~L.~Klimchitskaya,
 D.~E.~Krause, and V.~M.~Mostepanenko,
Phys. Rev. D {\bf 75}, 077101 (2007).
\bibitem{19}
F.~Chen,  G.~L.~Klimchitskaya,
V.\ M.\ Mos\-te\-pa\-nen\-ko, and U.~Mohideen,
Phys. Rev. B  {\bf 76}, 035338 (2007).
\bibitem{20}
E.~Buks and M.~L.~Roukes,
Phys. Rev. B {\bf 63}, 033402 (2001).
\bibitem{21}
V.~B.~Bezerra, R.~S.~Decca, E.\ Fischbach, B.\ Geyer,
G.\ L.\ Klimchitskaya, D.\ E.\ Krause, D.\ L\'opez,
V.\ M.\ Mostepanenko, and C.\ Romero,
Phys. Rev. E {\bf 73}, 028101 (2006).
\bibitem{22}
J.~S.~H{\o}ye, I.~Brevik, J.~B.~Aarseth, and K.~A.~Milton,
{J. Phys. A}: Math. Gen. {\bf 39}, 6031 (2006).
\bibitem{23}
J.~M.~Obrecht, R.~J.~Wild, M.~Antezza, L.~P.~Pitaevskii,
S.~Stringari, and E.~A.~Cornell,
Phys. Rev. Lett. {\bf 98}, 063201 (2007).
\bibitem{24}
M.~Antezza, L.~P.~Pitaevskii, and S.~Stringari,
Phys. Rev. A {\bf 70}, 053619 (2004).
\bibitem{27}
M.~Antezza, L.~P.~Pitaevskii, and S.~Stringari,
Phys. Rev. Lett. {\bf 95}, 113202 (2005).
\bibitem{25a}
S.~K.~Lamoreaux, { Phys. Rev. Lett.}
{\bf 78}, 5 (1997).
\bibitem{25}
R. S. Decca, E.
Fischbach, G. L. Klimchitskaya, D. E. Krause, D. L\'opez, and V.
M. Mostepanenko, Phys. Rev. D {\bf 68}, 116003 (2003).
\bibitem{26}
R.~S.~Decca, D.~L\'opez, E.~Fischbach, G.~L.~Klimchitskaya,
D.~E.~Krause, and V.~M.~Mostepanenko,
Ann. Phys. (N.Y.) {\bf 318}, 37 (2005).
\bibitem{29}
M.~Antezza,
J. Phys. A: Math. Gen. {\bf 39}, 6117 (2006).
\bibitem{30}
L.~P.~Pitaevskii,
J. Phys. A: Math. Gen. {\bf 39}, 6665 (2006).
\bibitem{32}
{\it Material Science and Engineering Handbook}, 3rd ed., edited by
J.\ F.\ Shackelford and W.\ Alexander
(CRC Press, Boca Raton, 2001).
\bibitem{33}
N.~P.~Bansal and R.~H.~Doremus,
{\it Handbook of Glass Properties}
(Academic Press, New York, 1986).
\end{thebibliography}
\end{document}